\documentclass{svjour3}                     
\smartqed  
\usepackage{graphicx}
\usepackage{natbib}
\usepackage{amsmath,amssymb,amsfonts}
\usepackage{mathptmx}      
\newcommand{\C}{\chi}
\journalname{J. Math. Biol.}
\begin{document}

\title{Bistability in two-locus models with selection, mutation, and recombination
}


\author{Su-Chan Park         \and
        Joachim Krug 
}


\institute{S.-C. Park \at
              Institut f\"ur Theoretische Physik, Universit\"at zu K\"oln, 50937 K\"oln, Germany \\
              Department of Physics, The Catholic University of Korea, Bucheon 420-743, Korea\\
              \email{spark0@catholic.ac.kr}           
           \and
           J. Krug \at
              Institut f\"ur Theoretische Physik, Universit\"at zu K\"oln, 50937 K\"oln, Germany \\
              \email{krug@thp.uni-koeln.de}           
}
\date{Received: date / Accepted: date}

\maketitle

\begin{abstract}
The evolutionary effect of recombination depends crucially on the
epistatic interactions between linked loci. A paradigmatic case where
recombination is known to be strongly disadvantageous is a two-locus
fitness landscape displaying reciprocal sign epistasis
with two fitness peaks of unequal height.
Although this type of model has been studied since the 
1960's, a full analytic understanding of the stationary
states of mutation-selection balance was not achieved so far.
Focusing on the bistability arising due to the recombination,
we consider here the deterministic, haploid two-locus model with reversible mutations,
selection and recombination. We find analytic formulae for the critical
recombination probability $r_c$ above which two stable stationary solutions
appear which are localized on each of the two fitness peaks.
We also derive the stationary genotype frequencies in various parameter
regimes. In particular, when the recombination rate is close to $r_c$ and the
fitness difference between the two peaks is small, we obtain a compact
description in terms of a cubic polynomial which is analogous to the Landau
theory of physical phase transitions.
\end{abstract}
\section{\label{Sec:Intro}Introduction}
After more than a century of research, the evolutionary basis of sex and recombination remains enigmatic \citep{Otto2009}.
In view of the complex evolutionary conditions faced by natural populations, the search for a single answer to the question of why 
sexual reproduction has evolved and is maintained in the vast majority of eukaryotic species may well be futile. Nevertheless, theoretical
population genetics has identified several simple, paradigmatic scenarios in which the conditions for an evolutionary advantage of sex
can be identified in quantitative terms, and which are therefore open (at least in principle) to experimental verification. 

One such scenario was proposed in the context of the adaptation of a 
population in a constant environment encoded by a fitness landscape, 
which assigns fitness values to all
possible genotypes. In this setting, the relative advantage of sexual
reproduction with respect to, say, the speed of adaptation or the mean
fitness level at mutation-selection balance, depends crucially on the
epistatic interactions between different loci
\citep{deVisser2007,Kouyos2007}. In its simplest form, epistasis is
associated with the curvature of the fitness surface, that is, the
deviation of the fitness effect of multiple mutations, all of which
are either deleterious or beneficial, 
compared to that predicted under the assumption of independent
mutations which contribute multiplicatively or additively to fitness.
It is well established that recombination
speeds up the adaptation in such a fitness landscape if 
the epistatic curvature is negative, in the sense that the fitness of multiple
mutants is lower than expected for independent
mutations \citep{Kondrashov1988}. 

\begin{figure}
\begin{center}
\mbox{
\includegraphics[width=0.7\textwidth]{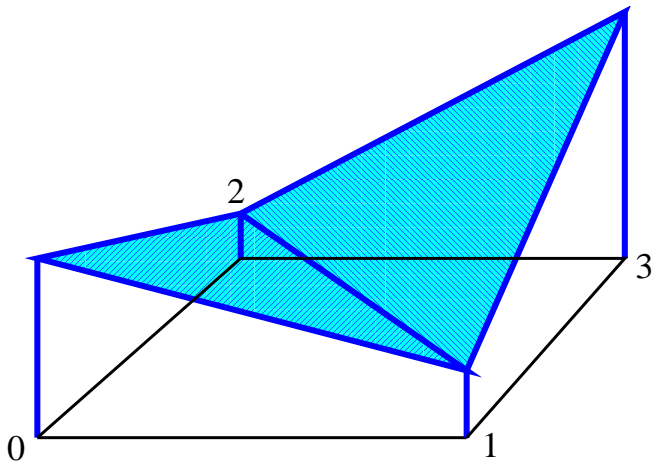}
}
\end{center}
\caption{\label{Fig:2peak} Schematic of a two-allele, two-locus fitness landscape
  with reciprocal sign epistasis and unequal peak heights. Fitness is
  represented by the height above the plane in which the four
  genotypes (labeled 0, 1, 2, 3) reside.}
\end{figure}

Unfortunately, experimental evidence indicates that negative epistasis is
not sufficiently widespread to provide a general explanation for the 
evolution of recombination
\citep{deVisser1997,Elena1997,Bonhoeffer2004}. 
Instead, empirical studies have highlighted the importance of a more complex
form of epistasis, termed \textit{sign epistasis}, in which not only
the magnitude but also the sign of the fitness effect of a given
mutation depends on the presence of mutations at other loci 
\citep{Weinreich2005}.  A simple example of sign epistasis is
shown in Fig.~\ref{Fig:2peak}, which depicts a haploid two-locus
fitness landscape. In this case the sign epistasis between the two loci
is reciprocal, which leads to the appearance of two fitness peaks
separated by a fitness valley \citep{Poelwijk2007}. 

In the two-locus
setting sign epistasis can
be viewed as an extreme case of positive epistasis, and it may be
expected to lead to a disadvantage of recombination. 
In a recent simulation study of the effect of
recombination on empirically motivated five-locus fitness landscapes displaying
sign epistasis \citep{deVisser2009}, we found that recombination can
hamper adaptation on such a landscape in a rather dramatic way:
Instead of moving to the global fitness optimum, populations get
trapped at local optima from which (in the limit of infinite
population size) they never escape. Mathematically, these numerical
calculations suggest the
appearance of multiple stable stationary solutions of the deterministic, 
infinite population dynamics above a threshold value $r_c$ of the 
recombination rate.

The simplest situation where this phenomenon
can occur is the haploid two-locus landscape shown in Fig.~\ref{Fig:2peak},
where it implies a bistability with two stationary solutions localized
near each of the fitness peaks labeled by 0 and 3. 
Indications for the occurrence of
bistability in this system can be found in several earlier studies of
the two-locus problem. \citet{CK1965} derived a condition
for the initial increase of the high peak mutant in a population
dominated by the low peak genotype.   
\citet{Eshel1970} established a sufficient condition for the low peak
population to remain trapped for all times when mutations are 
unidirectional ($0 \to 1,2 \to 3$), which was subsequently refined by
\citet{Karlin1971}. \citet{Feldman1971} and  
\citet{Rutschman1994} obtained conditions for the existence of multiple
stationary solutions in the absence of mutations.
The case of symmetric fitness peaks was
considered as a model for compensatory mutations in RNA secondary
structure by \citet{Stephan1996} and \citet{H1998}, who also addressed 
the role of genetic drift. Recent studies have considered the effect of recombination 
on the dynamics of peak escape in the asymmetric two-locus landscape for finite
as well as infinite populations \citep{Weinreich2005a,Jain2010}.
 
The corresponding diploid problem has also been studied extensively
\citep{Kimura1956,Bodmer1967}. \citet{B1989,Buerger2000} 
established conditions for bistability in a class of quantitative genetics models of stabilizing 
selection that are formally equivalent to the diploid version of our problem
with symmetric fitness peaks. 
Finally, bistability induced by recombination has
been observed in multilocus models in the context of quasispecies
theory \citep{Boerlijst1996,Jacobi2006} and evolutionary computation 
\citep{Wright2003}. 

However, a comprehensive analysis of the paradigmatic haploid two-locus system
with reversible mutations, reciprocal sign epistasis and fitness peaks
of unequal height does not seem to be
available, and the present paper aims to fill this gap.
In the next section we introduce the evolutionary dynamics
used in this work. We then show that finding the 
stationary solutions of the model amounts to analyzing the zeros of 
a fourth order polynomial, and devote the bulk of the paper to
extracting useful information about the critical recombination rate,
the genotype frequency distribution, 
and the mean fitness in various parameter regimes. 
We conclude with a summary of our results 
and a discussion of open problems.

\section{\label{Sec:Model}Model} 
We consider a haploid, diallelic two-locus system. 
The alleles at a locus are denoted by 0 and 1, 
and the four genotypes are labeled according to 
$00 \to 0, \; 01 \to 1, \; 10 \to 2, \; 11 \to 3$. 
Each genotype $i$ is endowed with a fitness $w_i$, which defines 
the fitness landscape. The population evolves in discrete,
non-overlapping generations under the influence of 
selection, mutation, and recombination. Since we are interested in the
emergence of bistability, in the sense of the existence of multiple stationary
frequency distributions, 
the limit of infinite population size is assumed. Thus, the 
frequency of each genotype evolves deterministically. The frequency changes 
according to the following order: selection, mutation, then recombination.

Let $f_i(\tau)$ denote the frequency of genotype $i$ at 
generation $\tau$. The frequency of the genotype at generation $\tau+1$ after
selection is proportional to the product of its frequency at generation 
$\tau$ and its fitness $w_i$.  After selection, mutations can 
change the frequency of genotypes.  We assume that the mutation probability 
per generation and locus, $\mu$, depends neither on the location of the 
locus in the sequence nor on the alleles, and mutations are 
assumed to be independent of each other.  Mathematically
speaking, the mutation from 
sequence $j$ to sequence $i$ then occurs with probability
\begin{equation}
U_{ij} = (1-\mu)^{2 - d(i,j)} \mu^{d(i,j)},
\end{equation}
where $d(i,j)$ is the Hamming distance between two sequences $i$ and $j$, i.e. the number
of loci at which the two sequences differ.  
After selection and mutation, the frequency distribution will be
\begin{equation}
\tilde f_i(\tau+1) = \sum_{j} U_{ij}
\frac{w_{j}}{\bar w\left (\tau \right )} f_{j}\left (\tau\right ),
\label{frequency_pool1}
\end{equation}
where $\bar w(\tau) \equiv \sum_i w_i f_i(\tau)$ is 
the mean fitness at generation $t$. 

After selection and mutation, recombination follows.  At first, 
two sequences, say $j$ and $k$, are selected with probability 
$\tilde f_{j} \tilde f_{k}$. With 
probability $1-r$, no recombination happens.  With probability $r$, however, 
the two sequences recombine in a suitable way and generate a recombinant, 
which may be identical to or different from $j$ and $k$ 
depending on the 
recombination scheme and the properties of the two initial sequences.  
One may interpret the case $r<1$ as a model for organisms which can
reproduce sexually as well as asexually. 
We will use the one-point crossover scheme, which is summarized as
\begin{equation}
\begin{array}{l}
\C_1 \C_2 \\
 \C_1'\C_2'
\end{array}
\rightarrow
\left \{
\begin{array}{lll}
\C_1&\C_2'&\quad \text{with prob. }\displaystyle r/2,\\
\C_1'&\C_2&\quad \text{with prob. }\displaystyle r/2,\\
\C_1&\C_2&\quad \text{with prob. }\displaystyle (1-r)/2,\\
\C_1'&\C_2'&\quad \text{with prob. }\displaystyle (1-r)/2,
\end{array}
\right .
\label{Eq:single_crossover}
\end{equation}
where $\chi_1$ ($\C_2$)  is the allelic type at the $1^\mathrm{st}$ ($2^\mathrm{nd}$) locus, 
$j = \C_1 \C_2$, $k = \C_1'\C_2'$ are parental genotypes,
and the four types on the right hand side are the possible resulting recombinants
with their respective probabilities. If $R_{i|jk}$ denotes the probability that
the resulting allelic type is $i$ in case types $j$ and $k$ (not necessarily 
different) attempt to recombine, the frequency of type $i$ at generation 
$\tau+1$ becomes $f_i' = \sum_{jk} R_{i|jk} \tilde f_j \tilde f_k$.

After selection, mutation, and recombination,
the frequency of each sequence in the next generation becomes
\begin{eqnarray}
\nonumber
\bar w(\tau) f_0' = 
f_0 w_0 (1-\mu)^2 + (f_1 w_1 + f_2 w_2) \mu (1-\mu) + f_3 w_3 \mu^2
- r(1-2\mu)^2 \frac{\delta(\tau)}{\bar w(\tau)},\\
\bar w(\tau) f_1' = 
f_1 w_1 (1-\mu)^2 + (f_0 w_0 + f_3 w_3) \mu (1-\mu) + f_2 w_2 \mu^2
+ r(1-2\mu)^2 \frac{\delta(\tau)}{\bar w(\tau)},
\nonumber\\
\bar w(\tau) f_2' = 
f_2 w_2 (1-\mu)^2 + (f_0 w_0 + f_3 w_3) \mu (1-\mu) + f_1 w_1 \mu^2
+ r(1-2\mu)^2 \frac{\delta(\tau)}{\bar w(\tau)},
\nonumber\\
\bar w(\tau) f_3' = 
f_3 w_3 (1-\mu)^2 + (f_1 w_1 + f_2 w_2) \mu (1-\mu) + f_0 w_0 \mu^2
- r(1-2\mu)^2 \frac{\delta(\tau)}{\bar w(\tau)},
\label{Eq:two_locus_eq}
\end{eqnarray}
where $\delta(\tau) = f_0(\tau) f_3(\tau) w_0 w_3 - f_1(\tau) f_2(\tau) 
w_1 w_2$ and $f_i$'s and $f_i'$'s mean the frequency at generation
$\tau$ and $\tau +1$, respectively.
Note that the case of free recombination frequently considered in the 
literature corresponds to $r=\frac{1}{2}$ in Eq.~(\ref{Eq:two_locus_eq}). 

By rearranging (\ref{Eq:two_locus_eq}), we get
\begin{eqnarray}
\frac{\bar w(\tau)}{ 1 - 2 \mu }\left ( f'_0- f'_3\right ) &=&   f_0w_0
- f_3w_3 ,
\label{Eq:pf1} \\
\frac{\bar w(\tau)}{ 1 - 2 \mu }\left ( f'_1- f'_2\right ) &=&   f_1w_1 - f_2w_2 ,\\
  f'_0 f'_3 - f'_1 f'_2 
&=&\frac{(1-r)(1-2\mu)^2}{ \bar w(\tau)^2}\delta(\tau).
\label{Eq:LD}
\end{eqnarray}
A nontrivial conclusion one can draw from Eq.~(\ref{Eq:LD}) is that 
linkage equilibrium, $f'_0 f'_3 = f'_1 f'_2$, is attained in one generation if $r=1$, which corresponds 
to the one-point crossover scheme with recombination probability 1. 

The stationary distribution is calculated by setting $f'_i=f_i$ in 
the above three equations, which gives
\begin{eqnarray}
\frac{f_0}{f_3} &=& \frac{\bar w -(1-2\mu) w_3}{ \bar w - (1-2\mu) w_0}\equiv A,\label{Eq:03}\\
\frac{f_1}{f_2} &=& \frac{\bar w -(1-2\mu) w_2}{ \bar w - (1-2\mu) w_1},\label{Eq:12}\\
\frac{f_0 f_3}{f_1 f_2} &=& \frac{ \bar w^2 - (1-r)(1-2\mu)^2 w_1 w_2}{ \bar w^2 - (1-r)(1-2\mu)^2 w_0 w_3}\equiv B,
\label{Eq:link}
\end{eqnarray}
where $\bar w$ is the mean fitness at stationarity.  With the two additional conditions 
\begin{equation}
\label{normalization1}
f_0 + f_1 + f_2 + f_3 = 1
\end{equation} 
and
\begin{equation}
\label{normalization2}
\bar w = w_0 f_0 + w_1 f_1 + w_2 f_2 + w_3 f_3 
\end{equation} 
the steady state solution is fully determined.  Without loss of generality, $w_3$ 
is set to be largest (with possible degeneracy).  For simplicity, we set 
$w_1=w_2$, which by (\ref{Eq:12}) implies that $f_1 = f_2 \equiv f$.  
Since the dynamics is invariant under multiplication of  
all fitnesses by the same factor, we can choose 
\begin{equation}
\label{fitnesses}
w_3=1, w_0=1-t, w_1=w_2=1-t-s
\end{equation}
with $0<t<1$ and $-t<s<1-t$.  For the sake of brevity, 
however, we will sometimes use $w_0$ and $w_1$ rather 
than $s$ and $t$ in what follows. The behavior for unequal valley fitnesses 
($w_1 \neq w_2$) will be discussed in Sect.\ref{Sec:asymmetric}. 

Note that the problem studied by 
\citet{H1998} corresponds 
to the case with $t=0$ and $0<s$. In this paper, we will assume that $t>0$ and 
$s>0$, that is, the fitness landscape has a unique global optimum and  
reciprocal sign epistasis (Fig.~\ref{Fig:2peak}).  
Unlike the fitness landscape with symmetric peaks, 
it is difficult to find the solution exactly (see also Appendix~\ref{Sec:Sympeak}). We will 
investigate the approximate solutions by assuming that some of the parameters are very small compared to others.

\section{\label{Sec:bi}Bistability}
\subsection{\label{Sec:gen}General behavior of solutions} 
Since the frequency of each genotype is strictly positive in the steady state, 
Eq. (\ref{Eq:03}) excludes the mean fitness $\bar w$ from being in the range between 
$(1-2\mu) w_0$ and $(1-2\mu) w_3$.  For obvious reasons, we will refer to a 
solution with $\bar w > (1-2\mu) w_3$ as high-fitness solution (HFS) and 
a solution with $\bar w<(1-2\mu) w_0$ as low-fitness solution (LFS).  
As an immediate consequence of Eq.~(\ref{Eq:03}), we see that 
a HFS (LFS) implies $f_3>f_0$ ($f_3<f_0$). 
The largest fitness being $w_3$, the existence of a HFS is expected regardless of the 
strength of the recombination probability (we will see later that this is 
indeed the case).  Hence we are mainly interested in the conditions which 
allow for a LFS.  Later, we will see that if it exists, there are actually two LFS's, 
only one of which is locally stable.  Since we are interested in  
stable solutions, in what follows LFS refers exclusively to the locally stable solution.  
Intuitively, the HFS should be locally stable, so the 
emergence of a LFS naturally implies bistability.

Without much effort, one can easily find necessary conditions for the 
bistability. First note that $\bar w$ is larger than $w_1$ by definition.  
Therefore, a LFS is possible only if $ w_1 < (1-2\mu) w_0$, or
\begin{equation}
\mu< \frac{s}{2(1-t)}.
\label{Eq:mu_cond}
\end{equation}
If we put $\mu =\frac{1}{2}$ in Eq.~(\ref{Eq:two_locus_eq}),
every $f_i'$ becomes $\frac{1}{4}$ regardless of $r$ and the $f_i$'s.
Since this is a unique equilibrium state for $\mu =\frac{1}{2}$, 
it could have been anticipated that bistability 
requires a restriction on $\mu$.  A necessary condition on $r$  
can be obtained from Eq.~(\ref{Eq:link}). While the numerator of the expression defining
$B$ is always positive, the denominator would be negative for the LFS if  
$w_0 - (1-r) w_3 < 0$. Hence a necessary condition for bistability is 
\begin{equation}
r > 1 - \frac{w_0}{w_3} = t,
\label{Eq:r_cond}
\end{equation}
which appears also in earlier works
\citep{CK1965,Eshel1970,Karlin1971,Feldman1971,Rutschman1994}.  
Most of the calculations 
in this paper are devoted to refining the conditions for bistability. To 
this end, we will reduce the five equations 
(\ref{Eq:03},\ref{Eq:12},\ref{Eq:link},\ref{normalization1},\ref{normalization2})
to a single equation for $\bar w$.

Equations (\ref{Eq:03}) and (\ref{Eq:link}) along with the normalization
(\ref{normalization1}) yield the relations
\begin{equation}
f = \frac{\sqrt{A}}{2 \sqrt{A} + \sqrt{B} (1+A)},\;\;
f_3 = \frac{\sqrt{B}}{2 \sqrt{A} + \sqrt{B} (1+A)},\;\;
 f_0 = A f_3,
\label{Eq:reduce}
\end{equation}
where we have used the fact that the $f_i$'s should be positive,
and the definition (\ref{normalization2}) of the mean fitness $\bar w$ implies that
\begin{equation}
2 \left ( \bar w - w_1 \right ) = \sqrt{\frac{B}{A}} \left ( w_3 + w_0 A - (1+A)\bar w
\right ).
\label{Eq:eqbarw}
\end{equation}
Since our main focus is on the LFS, it is convenient to define the auxiliary variable
$x$ through 
\begin{equation}
\label{xdef}
\bar w = (1-2\mu) (w_0-x),
\end{equation}
which implies that $x < -t$ and $x > 0$ for the HFS and the LFS, respectively. 
Note that $x$ is equivalent to the mean fitness and equilibria
will be found in terms of $x$.
We also note for future reference that with this reparametrization, 
$A$ and $B$ in Eqs.~(\ref{Eq:03}) and (\ref{Eq:link}) become
\begin{equation}
\label{AA}
A = 1+\frac{t}{x},\quad
B = 1+ (1-r) \frac{ w_0 w_3 - w_1^2}{(w_0-x)^2 - (1-r) w_0 w_3}
\end{equation}
Taking the square on both sides of 
Eq.~(\ref{Eq:eqbarw}) results in a quartic equation 
\begin{equation}
\label{quartic}
h(x) \equiv h_0(x) + r h_1(x) = 0,
\end{equation}
where the polynomials $h_0(x)$ and $h_1(x)$ are \textit{independent of $r$}
(see Appendix~\ref{Sec:Hx} and the Mathematica file in  
online supplement for explicit expressions).

We can draw some general conclusions concerning solutions of the quartic equation 
by evaluating $h(x)$ at selected points.  Since $h(x)$ is negative at $x=0$, 
$x=-t$, and at the point $x=x_1$ defined by $(1-2\mu)(w_0 -x_1) =  w_1$, and the coefficient of 
fourth order term is positive (see Appendix~\ref{Sec:Hx}), there always exist solutions 
in $x > x_1$ and $x<-t$ which correspond to $\bar w <w_1$ and 
$\bar w > (1-2\mu) w_3$, respectively. The meaningless solution, $\bar w<w_1$, 
has appeared because we took squares on both sides of Eq.~(\ref{Eq:eqbarw}).
In the Mathematica file in  online supplement, we show that 
$h(x)$ is positive when $x = 1- t - 1/(1-2\mu)$, or $\bar w = w_3 = 1$. This proves,
as anticipated, that the HFS with the mean fitness in the range
$(1-2\mu) w_3 < \bar w< w_3$ is present for all values of $r$. 

\begin{figure}[t]
\begin{center}
\mbox{
\includegraphics[width=0.9\textwidth]{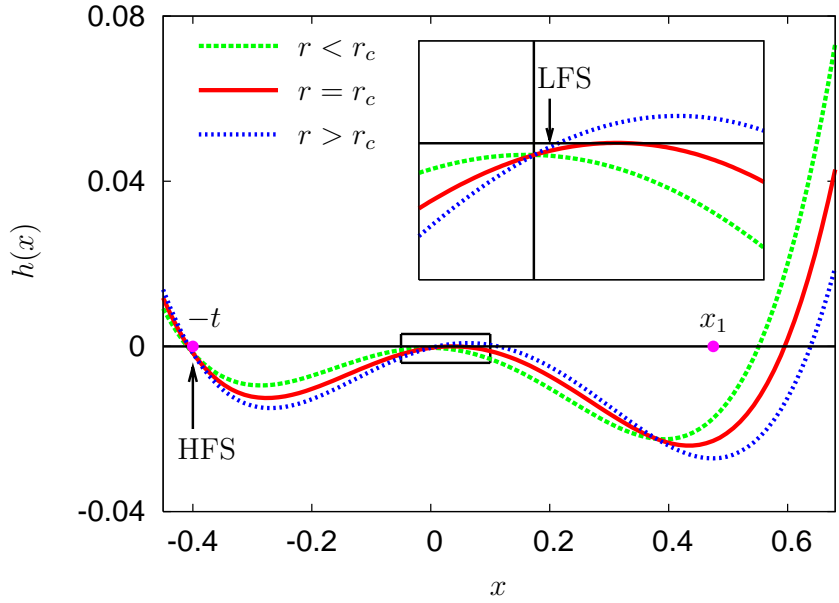}
}
\end{center}
\caption{\label{Fig:hx} Behavior of the function $h(x)$ 
around the critical recombination probability.  
In this example, $s=0.5$, $t=0.4$, and 
$\mu=0.01$ are used, which yields $r_c \approx 0.965$.  The curves meet at 
(three) points where $h_1(x)$ vanishes. 
The locations of the points $x=-t$ and $x=x_1$ are 
indicated by filled circles. 
Low  fitness solutions correspond to 
zero crossings in $0<x<x_1$, and high fitness solutions to those 
with $x<-t$. The HFS's are indicated by an arrow, which
happens to be close to $-t$.
The zero crossings with $x>x_1$ are spurious.
Inset: Close-up view of the boxed area. For 
clarity, the vertical line at $x=0$ is also drawn. One of the solutions of 
$h_1(x)=0$ happens to be close to $x=0$. Note that for $r=r_c$, two 
solutions become identical (degenerate). For $r>r_c$, the LFS is indicated by an arrow.}
\end{figure}

Hence the condition for bistability is recast 
as the condition for $h(x)$ to have a positive solution which is smaller than 
$x_1$. 
Because $h(0)$ and $h(x_1)$ are negative, the existence of a solution in the 
range $0<x<x_1$, or equivalently $w_1 < \bar w < (1-2\mu) w_0$, always entails two solutions in the same region if we count 
the number of degenerate solutions (that is, two identical solutions) as 2. Let us assume that  $h(x) = 0$ has 
two degenerate solutions  at $x = x_c$ when $r=r_c$ (as we will see, $r_c$ is the critical recombination rate above which bistability exists). This means 
that $x_c$ is the simultaneous solution of two equations $h(x_c) = h'(x_c)=0$, 
where the prime denotes the derivative with respect to the argument. Later, 
this simple relation will play a crucial role in finding $r_c$ as well as 
$x_c$.  Now let us change $r$ infinitesimally from $r_c$ to $r_c + \varepsilon_r$, and 
let the solutions of $h(x) = 0$ for $r=r_c + \varepsilon_r$ take the form 
$x_c + \varepsilon_x$. Note that we are only interested in solutions with 
$\varepsilon_x \rightarrow 0$ as $\varepsilon_r\rightarrow 0$ in the complex 
$x$ plane. Since two other solutions exist outside of the range $0<x<x_1$ and 
both $h(0)$ and $h(x_1)$ are negative, $h(x)$ with $r=r_c$ has local maximum 
at $x=x_c$, that is, $h''(x_c;r_c)<0$. Figure~\ref{Fig:hx} illustrates this 
situation. 

Using Eq.~(\ref{quartic}) we get
\begin{equation}
h(x_c + \varepsilon_x ;r_c + \varepsilon_r) \approx   
-\frac{1}{2} |h''(x_c;r_c) | \varepsilon_x^2 
+ \varepsilon_r h_1 (x_c) =0.
\label{Eq:dr}
\end{equation}
Hence real solutions are possible if the second term is positive. By definition, 
$r_c h_1(x_c)= - h_0(x_c)$. For $r=0$, Eq.~(\ref{quartic})
reduces to $h_0(x) = 0$, and we may conclude  
from the condition Eq.~(\ref{Eq:r_cond}), which is violated for $r=0$, that 
this equation does not have a solution in the 
region $0<x<x_1$. Hence $h_0(x_c)<0$ because $h_0(0)<0$ and $h_0(x_1)<0$ (see 
Appendix~\ref{Sec:Hx}), which, in turn, implies that $h_1(x_c)>0$. To summarize, we have proved that 
if $x_c$ is the degenerate solution of $h(x)=0$ when $r=r_c$, there are two solutions 
in the region $0<x<x_1$ when $r > r_c$. One should note that $r_c$, if it exists, is 
unique, as otherwise a contradiction to Eq.~(\ref{Eq:dr}) would arise. 

To establish the existence of bistability for $r>r_c$, it remains 
to find $r_c$. Even though general solutions of 
quartic equations are available, it is difficult to extract useful information 
from them.  Rather, we will look for approximate solutions by 
assuming that one of the three parameters $\mu$, $s$, and $t$ is much smaller 
than the other two. In fact, it follows from the condition (\ref{Eq:mu_cond}) that there cannot be any bistability
when $s \ll \mu$ (unless $t \to 1$). Hence, in this paper, we will not pursue 
the case where $s$ is the smallest parameter. 

Before turning to the derivation of the approximate solutions, we further
exploit the linear $r$-dependence of $h(x)$ in Eq.(\ref{quartic}). It implies that the
two equations $h(x_c;r_c) = h'(x_c;r_c) = 0$ with two unknowns can be reduced to a 
single equation for $x_c$ which does not involve $r_c$.  To be specific, from 
$h_0(x_c) + r_c h_1(x_c)=h_0'(x_c) + r_c h_1'(x_c)=0$ we obtain
\begin{equation}
r_c = - \frac{ h_0 (x_c)}{h_1(x_c)}
= - \frac{ h_0'(x_c)}{ h_1'(x_c)},
\label{Eq:rc_condition}
\end{equation}
or
\begin{equation}
 H(x_c) \equiv h_0 (x_c) h_1'(x_c) - h_1(x_c) h_0'(x_c) =0.
\label{Eq:xc_condition}
\end{equation}
Thus, rather than finding $r_c$ and $x_c$ simultaneously, we will first find 
$x_c$ by solving Eq.~(\ref{Eq:xc_condition}), which in turn, gives $r_c$ 
from Eq.~(\ref{Eq:rc_condition}). Equation (\ref{Eq:xc_condition}) will 
be analyzed below, after we have introduced one more useful concept.

\subsection{\label{Sec:muc}Critical mutation probability}

As evidenced by Eq.~(\ref{Eq:mu_cond}), a finite critical recombination
probability $r_c$ can exist only if $\mu$ is sufficiently small.
Mathematically, this implies that $r_c$ diverges as $\mu$ approaches 
a certain critical mutation probability $\mu_c$, such that bistability is not
possible for $\mu > \mu_c$. Although $r$ cannot, strictly speaking, exceed unity,
we will see later that this definition of $\mu_c$ will be of great use to find an accurate
expression for $r_c$. 
Setting $r_c=\infty$ in Eq.~(\ref{Eq:rc_condition}), we see that $\mu_c$  is the solution of 
the equations $h_1(x_c) = h_1'(x_c) = 0$.
Since $h_1(x)$ is a cubic function taking the form 
$h_1(x) =  -C_3 x^3 - C_2 x^2 + C_1 x - C_0$, $x_c$ is also the solution of the equation
\begin{equation}
G(x) = x h_1'(x) - 3 h_1(x)= C_2 x^2 - 2 C_1 x + 3 C_0 =0.
\end{equation}
From $G(x)$ and $h_1'(x)$, we can construct two linear equations for $x$ such that
\begin{eqnarray}
G_1(x) = C_2 h_1'(x) + 3 C_3 G(x) = - 2 (3 C_1 C_3 + C_2^2) x + C_1 C_2 
+ 9 C_0 C_3=0,\\
G_2(x) = \frac{1}{x} ( C_1 G(x) - 3 C_0 h_1'(x)) = (C_1 C_2 + 9 C_0 C_3) x
- 2 (C_1^2 - 3 C_0 C_2)=0,
\end{eqnarray}
where we have used the fact that $x_c \neq 0$. Hence, the value of 
$x_c$ for $r_c =\infty$ is given by
\begin{equation}
x_c^\infty = \frac{C_1 C_2 + 9 C_0 C_3}{2(C_2^2 + 3 C_1 C_3 )}
= \frac{2(C_1^2 - 3 C_0 C_2)}{C_1 C_2 + 9 C_0 C_3}.
\label{Eq:muc_xc}
\end{equation}
Note that $C_i$'s depend on $\mu$, $s$, and $t$, which means we have 
an equation for $\mu_c$ such that
\begin{equation}
(C_1 C_2 + 9 C_0 C_3)^2- 4 (C_1^2 - 3 C_0 C_2)(C_2^2 + 3 C_1 C_3)  =0,
\label{Eq:muc}
\end{equation}
which, in turn, provides $x_c^\infty$ by inserting $\mu_c$ in Eq.~(\ref{Eq:muc_xc}).
In fact, Eq.~(\ref{Eq:muc}) is equivalent to the discriminant of cubic
polynomials (see the Mathematica file in  online supplement). 

We now assume that $s,t\ll 1$, which also implies $\mu_c \ll 1$ by Eq.~(\ref{Eq:mu_cond}).
Then to leading order the  $C_i$'s become
\begin{equation}
C_3 = 2s + t,\; C_2 = (t+2\mu)(2s+t) - s^2,\;  
C_1 = t(s^2 - 2 \mu (2 s + t)),\; C_0 =\mu^2 t^2,
\end{equation}
and Eq.~(\ref{Eq:muc}) becomes (see the Mathematica file in online supplement)
\begin{equation}
3  s^{10} \nu^2(1-t) \left ( 1+\nu-z(2+\nu)\right )^2 M(z) = 0,
\label{Eq:muc_eq}
\end{equation}
where $z = \mu/s$, $\nu = t/s$, and $M(z)$ is a cubic polynomial with
\begin{eqnarray}
M(z) = 
32 ( \nu+2) z^3 - ( 13 \nu^2 + 48 \nu + 48) z^2 + 2 (2 \nu^3 + 7 \nu^2 + 9 \nu + 6) z - (1+\nu)^2.
\end{eqnarray}
Note that $z=(1+\nu)/(2+\nu)$ cannot be a solution because of 
Eq.~(\ref{Eq:mu_cond}). Hence the critical mutation probability is the 
solution of the equation $M(z) = 0$. As shown in the Mathematica file in online supplement, 
$M(z)$ is an increasing function of $z$, which, along with $M(0)<0$, allows 
only one positive real solution of $M(z)=0$ unless $\nu=0$. 
One can find the exact solution in the Mathematica file in online supplement, which is not suitable 
to present here. We will just present the asymptotic behavior of the solution 
for later purposes (see the Mathematica file in online supplement).  When $t \ll s$ ($\nu \ll 1$),
\begin{equation}
\mu_c = \frac{s}{4} \left [ 1 - 3 \left ( \frac{ t}{4 s} \right )^{2/3}
\right ] \textrm{ and } x_c^\infty = \left (\frac{s t^2}{16} \right )^{1/3},
\label{Eq:muc_smallt}
\end{equation}
and when $s \ll t$ ($\nu \gg 1$)
\begin{equation}
\mu_c \approx \frac{s^2}{4 t} \textrm{ and } x_c^\infty = \frac{s^2}{4 t}.
\label{Eq:muc_smalls}
\end{equation}
Although we derived the above two expressions from the exact solution 
(see the Mathematica file in online supplement), it is not difficult to find the asymptotic 
behavior without resorting to the exact solution.  When $\nu$ is finite, 
it would be more useful to have a numerical value. In the case $\nu=1$ ($t=s$) we get
\begin{equation}
\mu_c \approx 0.107 s \textrm{ and } x_c^\infty \approx 0.0616 s.
\label{Eq:muc_st}
\end{equation}
Below we will see how $\mu_c$ can be used to derive improved approximations for
$r_c$.
\subsection{\label{Sec:Smallmu}Approximation for small mutation probability} 
Now we will move on to finding the critical recombination probability.
We begin with the investigation of the approximate solutions 
for small mutation probability ($\mu \ll s, t$).
Let us assume that $x_c = x_0 + a_\mu \mu
+ O(\mu^2)$, which should be justified self-consistently.  
For later purposes, we introduce the
parameters
\begin{equation}
\label{alphabeta}
\alpha =  (1-t)(s+t)^2 - s^2, \;\;\;  \beta =  (1-t)(s+t)^2 + s^2. 
\end{equation}
Note that $\alpha$ is positive
if $s < \sqrt{1-t} + 1 -t$, which is automatically
satisfied because $s$ is smaller than  $1-t$ by definition. 

The leading behavior of $H(x_c)$ becomes
\begin{equation}
H(x_c) = x_0^2 (t+x_0)^2 ( a_2 x_0^2 + 2 a_1 x_0 + a_0) + O(\mu),
\end{equation}
where the $a_i$'s are parameters satisfying $a_1^2 - a_0 a_2 < 0$ 
(see the Mathematica file in online supplement). Hence, the solutions
for $x_0$ are 0, $-t$, and two complex numbers. Since 
$x_c$ must be positive, the only possible solution for $x_0$ 
is $x_0=0$. 
Accordingly, the actual leading behavior of $H(x_c)$ becomes (contributions of order $O(1)$ and $O(\mu)$ vanish)
\begin{equation}
H(x_c) = - \mu^2 s^2 t^2 \left ( (1-t)^2 \alpha - a_\mu^2 \beta \right ) 
+ O(\mu^3) = 0,
\end{equation}
which gives
\begin{equation}
a_\mu = (1-t) \left (\frac{\alpha}{\beta}\right )^{1/2}.
\end{equation}
By putting $x_c=a_\mu \mu$ into Eq.~(\ref{Eq:rc_condition}) and
keeping terms up to $O(\mu)$, we find 
\begin{equation}
r_c =  t + 2 \frac{1-t}{s^2} \left (
\alpha + \sqrt{\alpha \beta} \right ) \mu \equiv t + c_\mu \mu,
\label{Eq:rc_mu}
\end{equation}
which clearly satisfies the bound Eq.~(\ref{Eq:r_cond}),
and shows that this bound becomes an equality when $\mu \to 0$. 

The approximation for $r_c$ can be significantly improved 
by matching the approximation for small $\mu$ with 
the behavior of $r_c$ when $\mu$ is close to $\mu_c$. 
Since $r_c$ becomes infinite at $\mu = \mu_c$, 
we write
\begin{equation}
r_c =  t \frac{1 + \rho \mu}{1 - \mu/\mu_c},
\quad \textrm{ with }\quad 
\rho = 2 \frac{1-t}{s^2 t} \left ( \alpha + \sqrt{\alpha \beta} \right ) - 
\frac{1}{\mu_c}.
\label{Eq:rc_match}
\end{equation}
where $\rho$ is determined such that the correct leading behavior
is reproduced when $\mu$ is small. The specific form of the divergence at $\mu=\mu_c$ is motivated
by the behavior in the case of symmetric fitness peaks (see Eq.(\ref{Eq:rc0})). 

\begin{figure}
\includegraphics[width=\textwidth]{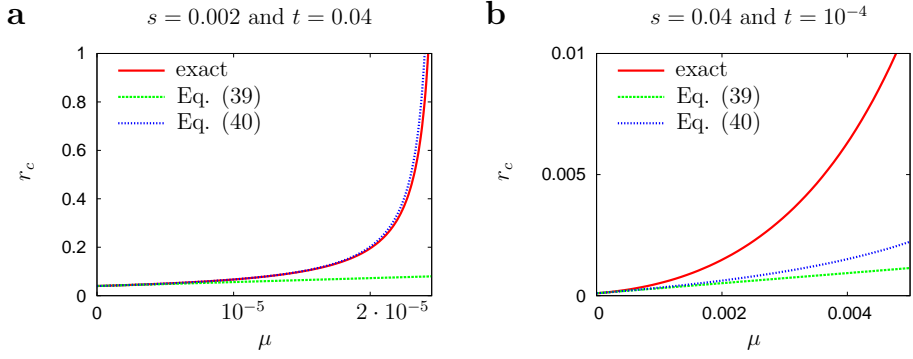}
\caption{\label{Fig:rcmu} Comparison of the exact $r_c$ with the
approximate solutions Eqs.~(\ref{Eq:rc_mu}) and (\ref{Eq:rc_match})
for three cases: (a) $s\ll t$ and  (b) $t\ll s$.
Eq.~(\ref{Eq:rc_match}) shows an excellent agreement with the
exact $r_c$ when $t$ is not too small compared to $s$. Eq.~(\ref{Eq:rc_mu})
is generally poor in predicting $r_c$.}
\end{figure}

In Fig.~\ref{Fig:rcmu}, the exact values of $r_c$ obtained from numerical 
calculations are compared with the two approximations Eqs.~(\ref{Eq:rc_mu}) 
and (\ref{Eq:rc_match}). One can find more such graphs in the Mathematica file in online supplement.
As expected, both expressions give a reliable
prediction when $\mu$ is sufficiently small. However, as the exact value of $r_c$
becomes larger, Eq.~(\ref{Eq:rc_mu}) does not give an accurate estimation,
which is not surprising. The more surprising observation is that
Eq.~(\ref{Eq:rc_match}) gives an excellent prediction for $r_c$ in 
almost all ranges. However, Eq.~(\ref{Eq:rc_match}) becomes a poor guidance 
for $r_c$ as $t$ gets smaller, which suggests that the case with small
$t$ should be treated separately.
In the next subsection, we will study the two-locus model for
small $t$.

\subsection{\label{Sec:Smallt}Approximation for small $t$} 
Now let us move on to the case with small $t$ ($t\ll \mu, s$) 
which connects the present study to the problem with symmetric 
fitness peaks considered by~\citet{H1998}. 
As before, we will find $x_c$ from Eq.~(\ref{Eq:xc_condition}). 

As shown by~\citet{H1998} (see also Appendix~\ref{Sec:Sympeak}), 
$x_c$ should approach zero as $t\rightarrow 0$. 
This can be rigorously shown from Eq.~(\ref{Eq:xc_condition}). 
If we assume that $x_c$ is finite as $t\rightarrow 0$, 
the leading order of Eq.~(\ref{Eq:xc_condition}) becomes
\begin{equation}
2s^2(2-s)(1-2 \mu)\left [\left \{(1-2\mu) x - 2(\mu_{c0} - \mu)\right \}^2 + 4 (1-s) \mu_{c0}^2 \right ] x^4 =0,
\end{equation}
where 
\begin{equation}
\label{Eq:muc0}
\mu_{c0} = \frac{s}{2(2-s)}
\end{equation} 
is the critical mutation
probability for the symmetric problem derived in Appendix~\ref{Sec:Sympeak}.
Since the terms in the square brackets are strictly positive, the only real solution is $x=0$.

It might seem natural to assume that $x_c = c_1 t+O(t^2)$ as in
Sec.~\ref{Sec:Smallmu}. However, this turns out to be wrong.
Not to be misled by an incorrect intuition, let us expand
$H(x_c)$ only with the assumption that $x_c$ is small, i.e., we do not specify
how small $x_c$ is compared to $t$.
Then, to leading order, the equation  $H(x_c)=0$ becomes
\begin{equation}
2 x_c^4 + 4 x_c^3 t - 2 a_t x_c t^2 - a_t t^3=0,
\label{Eq:t_app}
\end{equation}
where
\begin{equation}
a_t = \frac{(s-2\mu)^2 \mu^2}{2 s (1-2\mu)(2 \mu^2 + \mu_{c0} (s-4\mu))}.
\end{equation}
Note that terms of order $x_c^5$ and $x_c^6$ are neglected compared to $x_c^4$.
Actually, there is a term of order $x_c^2$ in $H(x_c)$, but its coefficient is $O(t^2)$, so
it is neglected compared to $x_c t^2$.
Let us assume that $x_c \sim t^\zeta$. If $\zeta \ge 1$, the solution
of Eq.~(\ref{Eq:t_app}) is $x_c = - t/2$ which does not fall into 
the range $0<x_c < x_1$.
If $\zeta < 1$, the leading behavior of Eq.~(\ref{Eq:t_app}) should be
$x_c^4 - a_t t^2 x_c = 0$ which gives
$x_c \approx (a_t t^2)^{1/3}$.
A more accurate estimate of $x_c$ is derived in the Mathematica file in online supplement,
which reads
\begin{equation}
x_c \approx (a_t t^2)^{1/3}-\frac{t}{2}.
\label{Eq:t_xc}
\end{equation} 
Note that the power $\frac{2}{3}$ was already observed when we 
calculated $\mu_c$ for $t\ll s$ in Eq.~(\ref{Eq:muc_smallt}).
From Eq.~(\ref{Eq:rc_condition}) along with Eq.~(\ref{Eq:t_xc}) we get
\begin{eqnarray}
r_c \approx r_{c0} \left ( 1 + \frac{3\mu_{c0} (2 \mu^2 + \mu_{c0} (s-4 \mu))}{2 s \mu^2 (\mu_{c0}-\mu)}
\left ( a_t t^2  \right )^{1/3} - \frac{2 \mu_{c0}^2 (1+s)}{s^2(\mu_{c0}-\mu)} t\right ),
\label{Eq:rc_t}
\end{eqnarray}
where $r_{c0}$ is the critical recombination for
$t=0$ given by (see Appendix~\ref{Sec:Sympeak} for derivation) 
\begin{equation}
r_{c0} = \frac{2 \mu^2}{(1-2\mu)(\mu_{c0} - \mu)}.
\label{Eq:rc0}
\end{equation}
We will use the same technique as in Sec.~\ref{Sec:Smallmu} to improve
the quality of the approximation for $r_c$. Since $r_c$ diverges
at $\mu = \mu_c$ rather than at $\mu_{c0}$ (note that $\mu_{c0} > \mu_c$),
we can associate the behavior for small $r_c$ with that for larger $r_c$
in such a way that
\begin{equation}
r_c = \frac{2 \mu^2}{(1-2\mu)(\mu_c - \mu)} ( 1  + \tilde \rho_0 t^{2/3} 
+ \tilde \rho_1 t),
\label{Eq:rc_t_match}
\end{equation}
where the coefficients $\tilde \rho_0$ and $\tilde \rho_1$ are
determined by requiring that
the leading behavior of $r_c$ in Eq.~(\ref{Eq:rc_t_match}) is same
as that in Eq.~(\ref{Eq:rc_t}). 
This yields
\begin{eqnarray}
\tilde \rho_0 = \frac{3 \mu_{c0}}{2 s(\mu_{c0}-\mu)}
\left \{ \frac{2 \mu^2 + \mu_{c0} (s- 4\mu)}{\mu^2} a_t^{1/3}    -(1-s)(2\mu_{c0})^{1/3} \right \},
\quad \tilde \rho_1 =0,
\end{eqnarray}
where we have used
a more accurate expression for $\mu_c$ than Eq.~(\ref{Eq:muc_smallt}), derived in the Mathematica file in online supplement,
which reads
\begin{equation}
\mu_c = \mu_{c0} - \frac{3(1-s)}{4(2-s)} (2 \mu_{c0} t^2)^{1/3}
+ \frac{2\mu_{c0}^2(1+s)}{s^2} t.
\end{equation}
For completeness, we present the corresponding $x_c^\infty$ which reads
\begin{equation}
x_c^\infty = \left ( \frac{\mu_{c0} t^2}{4} \right )^{1/3} - \frac{t}{2}.
\end{equation}

\begin{figure}
\includegraphics[width=\textwidth]{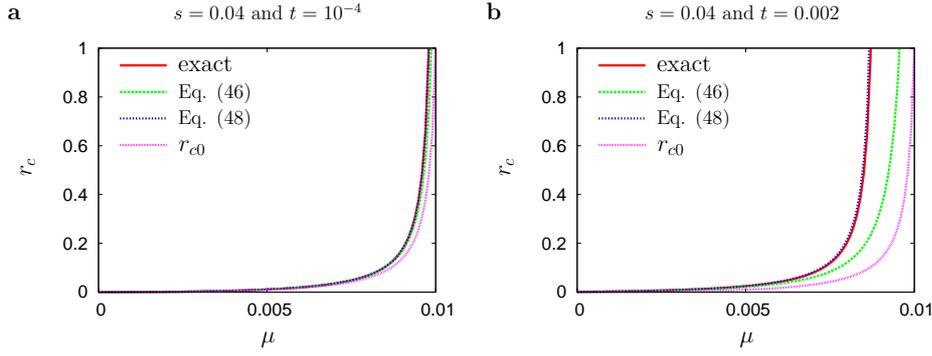}
\caption{\label{Fig:rct} Comparison of the exact $r_c$ with the
approximate solutions Eqs.~(\ref{Eq:rc_t}) and (\ref{Eq:rc_t_match})
for $s=0.04$ and (a) $t=10^{-4}$,  (b) $t=0.002$. 
For comparison, we also draw the critical recombination probability
when $t=0$ ($r_{c0}$).
For $t=10^{-4}$, both approximations are in good agreement with
the exact solutions. As $t$ increases, Eq.~(\ref{Eq:rc_t}) starts 
to deviate from the exact solution, but the improved approximation still
has predictive power.  }
\end{figure}

Figure~\ref{Fig:rct} compares the exact values with the approximations
Eqs.~(\ref{Eq:rc_t}) and (\ref{Eq:rc_t_match}). $s$ and $t$ in 
Fig.~\ref{Fig:rct}a are same as those in 
Fig.~\ref{Fig:rcmu}b. For sufficiently small $t$, both approximations
show a good agreement. As anticipated, Eq.~(\ref{Eq:rc_t}) becomes
worse as $t$ increases, even though Eq.~(\ref{Eq:rc_t_match}) is still in
good agreement with the exact solution. Needless to say,
Eq.~(\ref{Eq:rc_t_match}) fails 
when $t$ is not much smaller than $s$. Although Fig.~\ref{Fig:rct}
seems to suggest that the agreement is good even for very small $\mu$,
this is an artifact because $r_c$ itself is too small. In this regime,
one should use the approximation developed in Sec.~\ref{Sec:Smallmu}.

To summarize our findings up to now, 
we have provided approximate expressions for $r_c$ which are valid
in the specified regimes. Taken together, these
expressions cover essentially the full range of biologically relevant parameters.

\subsection{\label{Sec:Fre}Frequency distributions}
So far we have investigated the critical recombination and mutation
probabilities. To complete the analysis, we need to determine
the frequency distribution at stationarity.
For the LFS, this can be readily done using Eq.~(\ref{Eq:dr}).
From Eq.~(\ref{xdef}) we see that the solution with the smaller $x>0$
will confer the larger mean fitness. Let $x_s$ ($x_l$) be the smaller
(larger) positive solution. Equation~(\ref{AA}) shows that $B(x_s) < B(x_l)$
and $A(x_s) > A(x_l)$. Note that we are treating $A$ and $B$ as functions of
$x$. If we rewrite $f_3$ as $1/(2 \sqrt{A/B} + 1 + A)$, 
it is clear that $f_3(x_s)$ must be smaller than $f_3(x_l)$.
Introducing $\Delta f_3 
= f_3 (x_l) - f_3(x_s)$ and $\Delta f_0 = f_0(x_l) - f_0(x_s)$,
the mean fitness difference can be written as
\begin{equation}
\Delta \bar w \equiv \bar w(x_2) - \bar w(x_1)
= (w_0 - w_1)\Delta f_0  + (w_3 - w_1)\Delta f_3 .
\end{equation}
Since $\Delta \bar w$ is negative and $\Delta f_3$ is positive, 
$\Delta f_0$ must be negative. That is, the solution with the smaller 
$x>0$ should confer the larger value of $f_0$.
Below we will argue that the solution with the larger
$f_0$ is stable. Hence we limit ourselves to the study of the solution
with smaller $x$. We also present approximate expressions for the HFS.

As before, we have to conduct separate analyses depending
on which parameter is the smallest. 
For these analyses we will use the expansions 
Eqs.~(\ref{Eq:rc_mu}) and 
(\ref{Eq:rc_t}) for $r_c$ rather than the improved approximations 
Eqs.~(\ref{Eq:rc_match}) and (\ref{Eq:rc_t_match}), which are not
suited for a systematic perturbative solution.

We begin with the case when $\mu \ll s,t$.
The functions $h''(x_c,r_c)$ and $h_1(x_c)$ in (\ref{Eq:dr}) then become
\begin{equation}
- h''(x_c,r_c) \approx  2 t \beta,\;
h_1(x_c) \approx a_\mu s^2 t \mu,
\end{equation}
which gives
\begin{equation}
\varepsilon_x \approx -   x_c
\left (\frac{s^2 \varepsilon_r }{ (1-t)\sqrt{\alpha\beta} \mu}\right)^{1/2}
\equiv - x_c \epsilon,
\end{equation}
where $x_c = a_\mu \mu$ and the definition of $\epsilon$ is clear.
If we set $x = x_c+ \varepsilon_x$ and 
$r =  r_c + \varepsilon_r = t+ c_\mu \mu + \varepsilon_r$ from (\ref{Eq:rc_mu}), we get
\begin{equation}
A = 1 + \frac{t}{x} \approx \frac{t}{x_c(1 - \epsilon)},\quad
B = \frac{\alpha}{t \left (\varepsilon_r - 2 \varepsilon_x + (c_\mu - 2 a_\mu) \mu\right )},
\end{equation}
which are large.
Note that $A$ becomes negative when $\epsilon> 1$, which means that
the regime of validity of the above approximation is quite narrow.
To leading order, the population frequencies are then obtained from
(\ref{Eq:reduce}) as
\begin{eqnarray}
f \approx \frac{1}{ \sqrt{A B}} \approx 
 \sqrt{\alpha}\frac{1-t}{s}\mu
\left (1 + \sqrt{\frac{\alpha}{\beta} } ( 1- \epsilon)\right ) 
,\;
f_3 \approx \frac{1}{A} \approx \frac{1-t}{t} \sqrt{\frac{\alpha}{\beta}}(1 - \epsilon)\mu,
\label{Eq:small_mu_f0}
\end{eqnarray}
and, by normalization, $f_0 = 1 - f_3 - 2 f$. The second (unstable) solution is 
obtained by setting $\epsilon \mapsto - \epsilon$.

To find the HFS, we shift the variable $x$ 
to $y = -(x+t)$ and look for a solution of $g(y) \equiv h(-t -y)=0$.
As shown in the Mathematica file in online supplement, the HFS is located at $y = 0$ ($x = -t$) 
for $\mu \rightarrow 0$. This is a consequence
of the fact that when $\mu=0$, the stationary fitness of the HFS 
is $\bar w = w_3$. If we now set $y = \sum_{n\ge 1} y_n \mu^n$ and
expand $g(y)$ as a power series in $\mu$, the equation $g(y)=0$
gives (see  the Mathematica file in online supplement)
\begin{equation}
y_1 = 0,\; y_2 = \frac{r(1-s-t)^2 +  (s+t)(2 - s - t)}{(s+t)^2(r(1-t)+t)} t,\;
y_3 = 2 y_2 \frac{ t(s+t)^2 (1-r) + r (2 s + t)}{(r (1-t) + t)(s+t)^2}
\label{Eq:y2}
\end{equation}
Up to $O(\mu^2)$, the genotype frequencies for the HFS become
\begin{equation}
f \approx  \frac{\mu}{s+t}(1-y_2 \mu),\quad
f_0 \approx A \approx \frac{y_2}{t} \mu^2,
\label{Eq:small_mu_f3}
\end{equation}
and $f_3 = 1 - 2 f - f_0$. 

One can see qualitative differences between Eqs.~(\ref{Eq:small_mu_f3})
and (\ref{Eq:small_mu_f0}). 
First, the frequency of the less populated fitness peak genotype is different in the two cases. 
In Eq.~(\ref{Eq:small_mu_f0}), $f_3 = O(\mu)$, 
but in Eq.~(\ref{Eq:small_mu_f3}), $f_0 = O(\mu^2)$.
However, this tendency cannot persist when $r$ is large. For example, if
$r=1$, Eq.~(\ref{Eq:link}) suggests that $f_3$ should be $O(\mu^2)$
provided $f$ is still $O(\mu)$. Hence, this qualitative difference only occurs
when $r$ is close to $r_c$.
Second, the leading behavior of the frequency $f$ of valley genotypes
does not depend on $r$ in Eq.~(\ref{Eq:small_mu_f3}), 
which is not true in Eq.~(\ref{Eq:small_mu_f0}) because of the
dependence on $\epsilon$.

To analyze the stability of the solutions, we linearize Eq.~(\ref{Eq:two_locus_eq})
at the steady state frequency. For the stability analysis, we assume that 
$f_1(\tau) = f_2(\tau)$ for all $\tau$, which is true if they are equal
initially. The linearization then yields a square matrix with rank 2, whose
largest eigenvalue (in absolute value) determines the stability. 
For the HFS, the eigenvalues up to $O(\mu^0)$ are
$1-s-t$ and $(1-t)(1-r)$ which are smaller than 1. Hence, the HFS is
always stable. At $r=r_c$, the largest eigenvalue for the LFS is expected
to be 1. Since we are restricted to an approximation up to $O(\mu)$, all we can show
is that the largest eigenvalue of the LFS is $1+O(\mu^2)$ at $r=r_c$. 
In the Online Supplement,
we show that the largest eigenvalue for $s=t$ becomes
$1+O(\mu^2)$ and the smaller one is $(1-s-t)/(1-t)$. When treated
numerically, it is easy to see that the stable solution indeed corresponds
to the smaller $x$ (details not shown).

The next step we will take is to find the frequency distribution of
the LFS in the case that $t$ is much smaller than $s$ and $\mu$.
From Eqs.~(\ref{Eq:t_xc}) and (\ref{Eq:rc_t}), we get (up to
leading order)
\begin{eqnarray}
- h''(x_c,r_c) = \frac{6 s (2 \mu^2 + \mu_{c0} (s-4\mu))}{\mu_{c0} -\mu} (a_t t^2)^{1/3},\nonumber\\
h_1(x_c) = 2s(2-s)(1-2\mu)(\mu_{c0}-\mu) (a_t t^2)^{2/3},
\end{eqnarray}
thus from Eq.(\ref{Eq:dr})
\begin{equation}
\varepsilon_x =- (\mu_{c0}-\mu)(a_t t^2)^{1/3} 
\left ( \frac{2 (2-s)(1-2\mu) }{3(2 \mu^2 + \mu_{c0} (s-4\mu))} 
\frac{\varepsilon_r}{(a_t t^2)^{1/3}}
\right)^{1/2}\equiv - x_c \eta,
\end{equation}
where we have kept $x_c = (a_t t^2)^{1/3}$ up to leading order from Eq.~(\ref{Eq:t_xc}) and $\eta$ has an obvious meaning.
Accordingly, $A$ and $B$ become
\begin{eqnarray}
A &\approx& 1 + A_t,\\
\sqrt{B}&\approx& \frac{s-2\mu}{2 \mu} \left ( 1 + B_t x_c + B_r \varepsilon_r
+ B_x \varepsilon_x \right ),
\end{eqnarray}
where
\begin{eqnarray}
A_t = \frac{t}{x_c} \left ( 1  - \eta\right )^{-1},\quad
B_t = \frac{8 \mu^2 - 4 s \mu (1+\mu) - s^2(1-4\mu)}{8 a_t (s^2 + 8 \mu^2
-4 s \mu (1+\mu))},\nonumber \\
B_r = - \frac{2 s^2 \mu^2}{\mu_{c0} (s-2\mu)^2 r_{c0}^2} ,\quad
B_x = \frac{s(1-2\mu)(s-4\mu)(\mu_{c0} - \mu)}{2 (s-2\mu)^2 \mu^2}.
\end{eqnarray}
The above approximation is valid only when $\eta\ll 1$ ($\varepsilon_r \ll
t^{2/3}$).
Note that unlike the previous case, $A$ is close to $1$ ($A_t \sim t^{1/3}$).
Hence the frequency distribution for the LFS becomes
\begin{eqnarray}
f \approx \frac{\mu}{s} \left ( 1 - 2 \left (\frac{1}{2}-\frac{ \mu}{s}
\right )\left (  B_t x_c + B_r \varepsilon_r +B_x\varepsilon_x + \frac{A_t^2}{8}\right )\right )
\label{Eq:f1t}
,\\
\label{Eq:f3t}
f_3 \approx \left (\frac{1}{2} - \frac{\mu}{s} \right ) 
\left (1 - \frac{A_t}{2} 
\right ) ,\\
f_0 \approx \left (\frac{1}{2} - \frac{\mu}{s} \right ) 
\left (1 + \frac{A_t}{2} 
\right ),
\label{Eq:f0t}
\end{eqnarray}
where we have kept the leading order of each term.
Since the above approximation requires that
$\varepsilon_r = r-r_c \ll t^{2/3}$, it
cannot reproduce the symmetric solution in Appendix~\ref{Sec:Sympeak}, which applies
when $t \to 0$ at fixed $r$.

\subsection{\label{Sec:Landau}Landau theory} 
In this subsection, we develop an 
approximation that is valid when $r$ is close to $r_c$ and the
asymmetry of the fitness landscape is small, in the sense that $t$ is
smaller than all other parameters. This approximation is inspired by
the Landau theory from the physics of phase transitions, and it will
allow us to represent both the LFS and the HFS in a simple, compact
form.
 
We start from the observation that, according to 
Eqs.~(\ref{Eq:f1t}), (\ref{Eq:f3t}), and (\ref{Eq:f0t}), the
valley genotype frequency $f \approx \mu/s$ in the regime of interest,
with the peak frequencies $f_3$ and $f_0$ symmetrically placed around
$1/2 - \mu/s \approx 1/2 - f$. Moreover, the difference $f_0 - f_3
\approx A_t \sim t^{1/3}$ becomes small for $t \to 0$.  
This motivates the parametrization
\begin{equation}
f_0 = \left ( \frac{1}{2} - f \right )
\left ( 1 - u \right ),
f_3 = \left ( \frac{1}{2} - f \right )
\left ( 1 + u \right )
\label{Eq:defu}
\end{equation}
which defines the new variable $u$.
Inserting this into Eq.~(\ref{Eq:pf1}) with $f'_i = f_i$ we obtain
\begin{equation}
\bar w = (1-2\mu) \left ( 1 + \left ( 1 - u \right ) \frac{t}{2 u} \right ).
\label{Eq:barwlandau}
\end{equation}
On the other hand, from the definition (\ref{normalization2}) of $\bar w$, 
we find a relation between $f$ and $u$ such that
\begin{equation}
f = - \frac{t}{2 u} \frac{(1-u)(1+u-2\mu)}{2 s + t(1+  u)} +
\frac{2\mu}{2 s + t (1+u)}.
\label{Eq:fu}
\end{equation}
Up to now, everything is exact.
Note that when $u\ll 1$ and $t\ll u$, the leading behavior of
Eq.~(\ref{Eq:fu}) is $\mu/s$ which is consistent with the LFS frequency distribution
in Eq.~(\ref{Eq:f1t}). Moreover, as the mean
fitness for the HFS in the case of
small $t$ is not expected to deviate much from $1-2\mu$, the HFS also requires
that $t \ll u$. So for all solutions, the leading behavior of $f$
is $\mu/s$. This is rather different from the case when $\mu$ is the
smallest parameter.

By keeping the leading terms under the assumption that
$t \ll \mu \ll s \ll 1$ and $t\ll u \ll 1$,
from Eqs.~(\ref{Eq:LD}) and (\ref{Eq:fu}) we obtain the equation
\begin{equation}
t  -  ( r_0 - r  )u -  r  u^3
=0
\label{Eq:Landau}
\end{equation}
for $u$, where $r_0 = 8\mu^2/s$.
If we interpret $r$ as the (inverse) temperature, $t$ as the external 
magnetic field, and $u$ as the magnetization, this has 
precisely the form of the
Landau equation for the  para- to ferromagnetic phase transition
\citep{Plischke2006}.

The general solution of Eq.~(\ref{Eq:Landau}) can be written in a compact form.
Let 
\begin{equation}
\label{Eq:Delta}
\Delta = \left(\frac{t}{2 r} \right)^2 - \left( \frac{r-r_{0}}{3 r}
\right)^3
\end{equation}
denote the discriminant of Eq.~(\ref{Eq:Landau}).
When $\Delta>0$, there is only one real solution which reads
\begin{equation}
u_\mathrm{HFS} = 
\left ( \frac{t}{2 r} + \sqrt{\Delta}\right)^{1/3}+
\left ( \frac{t}{2 r} - \sqrt{\Delta}\right)^{1/3}.
\label{Eq:plusD}
\end{equation}
For $r$ sufficiently far below $r_0$, in the sense that 
$r_0 - r \gg  (t^2 r)^{2/3}$, this reduces to 
\begin{equation}
\label{Eq:u_HFS_below}
u_\mathrm{HFS} \approx \frac{t}{r_0 - r}, 
\end{equation}
which is the solution of Eq.~(\ref{Eq:Landau}) with the cubic term omitted. 
When $\Delta<0$, there are three real solutions
\begin{eqnarray}
u_\mathrm{HFS} =  2 \left ( \frac{r-r_{0}}{3 r} \right )^{1/2} 
\cos\frac{\theta}{3},\quad
u = -2 \left ( \frac{r-r_{0}}{3 r} \right )^{1/2}
\sin\left ( \frac{\pi}{6}\mp \frac{\theta}{3}\right ),
\label{Eq:minusD}
\end{eqnarray}
where $\tan\theta = 2 r \sqrt{|\Delta|}/t$ with $0\le \theta \le \pi/2$.
The stable LFS corresponds to the smallest value of $u$, which yields the larger $f_0$ among the two solutions with negative $u$  
(see the discussion in the beginning of Sect.~\ref{Sec:Fre}),
\begin{equation}
\label{Eq:Landau_LFS}
u_\mathrm{LFS} = - 2\left ( \frac{r-r_{0}}{3 r} \right )^{1/2}
\sin\left ( \frac{\pi}{6} + \frac{\theta}{3}\right ).
\end{equation}
One can easily see that for $t\rightarrow 0$ ($\theta\rightarrow
\pi/2$) the solutions (\ref{Eq:minusD}) and (\ref{Eq:Landau_LFS}) approach
the symmetric peak solutions
\begin{equation}
\label{Eq:symmetric_peaks}
u_\mathrm{HFS} = \sqrt{1 - r_0/r}, \;\;\; u_\mathrm{LFS} = - \sqrt{1 - r_0/r}. 
\end{equation}
The critical recombination probability can be found from $\Delta=0$
which gives
\begin{equation}
r_c = 8 \frac{\mu^2}{s} \left (
1 + \frac{3}{4} \left ( \frac{ st}{2 \mu^2} \right )^{2/3} 
+ O(t^{4/3})
\right ).
\end{equation}
Note that this agrees with 
Eq.~(\ref{Eq:rc_t}) only up to order $t^{2/3}$.

\begin{figure}
\begin{center}
\mbox{
\includegraphics[width=0.8\textwidth]{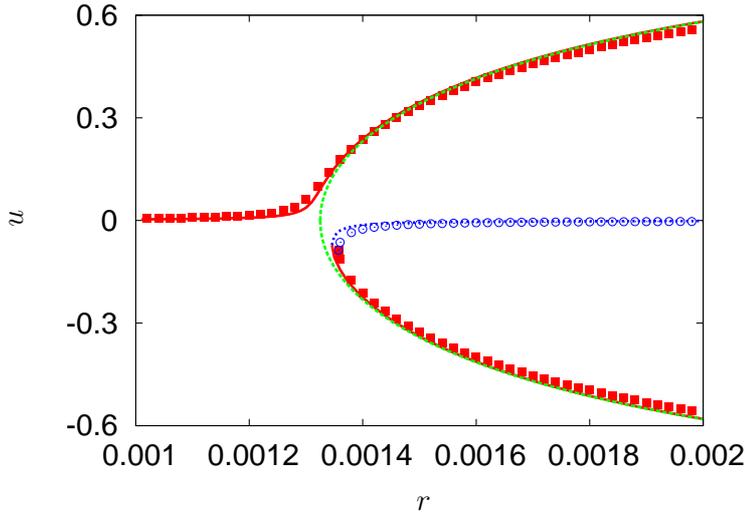}
}
\end{center}
\caption{\label{Fig:Landau} Plots of $u = (f_3 - f_0)/(1-2f)$ obtained from the
exact numerical solutions (symbols) and from Eq.~(\ref{Eq:Landau})
(full curves)
as a function of $r$ for $t=10^{-6}$, $s=10^{-2}$, and $\mu =
10^{-3}$. For these parameters $r_{0} = 8 \times 10^{-4}$ and $r_{c0}
\approx 1.325 \times 10^{-3}$. 
The filled squares are stable solutions and open circles are unstable
solutions. The dashed line shows the symmetric peak solutions
(\ref{Eq:symmetric_peaks}) for $r > r_{c0}$. 
The approximate solution is seen to be valid well beyond the regime where $\eta<1$.
}
\end{figure}

Although the LFS in Eqs.~(\ref{Eq:f1t}), (\ref{Eq:f3t}), and 
(\ref{Eq:f0t}) is valid only when $\epsilon_r \ll t^{2/3}$,
the approximate solutions Eqs.~(\ref{Eq:plusD}) and (\ref{Eq:minusD}) turn
out to be in good agreement with the exact solutions, provided $r_0$
is replaced by the exact critical recombination rate $r_{c0}$ for the
symmetric peak problem [see Eq.(\ref{Eq:rc0})].
In Fig.~\ref{Fig:Landau}, we compare the exact values of $u$ with
the approximate solutions for $t=10^{-6}$, $s=10^{-2}$, and $\mu = 10^{-3}$.
For these parameters, $\eta$ becomes larger than 1 when $\varepsilon_r \approx
3\times 10^{-5}$.

\subsection{\label{Sec:fitness}Behavior of the mean fitness}

\begin{figure}[t]
\includegraphics[width=\textwidth]{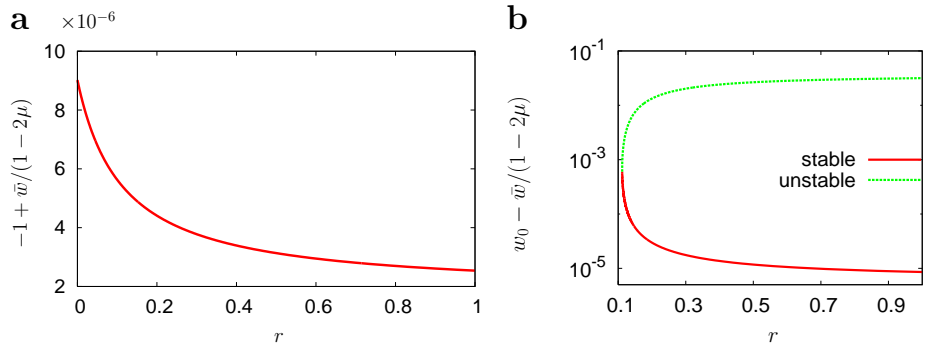}

\caption{\label{Fig:fitness}Behavior of mean fitness as a function of $r$ for 
(a) the HFS in linear scales and (b) the LFS in semi-logarithmic scales.  
The parameters are $t=s=0.1$ and $\mu = 0.001$.
For the purpose of clarity, we show the difference between $\bar w$ and its limiting value for
large $r$, which is $w_3(1-2 \mu)$ for the HFS and $w_0(1-2\mu)$ for the LFS.
(a) Mean fitness decreases with $r$.
(b) Mean fitness increases (decrease) with $r$ for the stable (unstable)
LFS. Note that this figure displays the positive 
quantity $w_0 - \bar w/(1 - 2 \mu)$.
}
\end{figure}

We are now prepared to discuss the
dependence of the mean population fitness $\bar w$ on the
recombination rate. Since mean fitness is linearly related to the
auxiliary variable $x$ through (\ref{xdef}), this amounts to examining how
the solutions of (\ref{quartic}) vary with $r$. Solving (\ref{quartic}) for $r$ we obtain   
\begin{equation}
r(x) = - \frac{h_0(x)}{h_1(x)}
\end{equation}
and $r'(x) = H(x) /h_1(x)^2$, where $H(x)$ was defined in (\ref{Eq:xc_condition}).
Since $H(x)$ has a unique root $x_c$ in the regime of
interest, we conclude that $r(x)$ displays a minimum at $x=x_c$.
Recalling that the stable LFS corresponds to the smaller of the two
solutions of (\ref{quartic}) with $x > 0$, it follows that 
the mean fitness of the stable (unstable) LFS increases (decreases)
with $r$. In addition, the fitness of the HFS must be a monotonic
function of $r$. The behavior of all three solutions is illustrated in 
Fig.\ref{Fig:fitness}, which shows that fitness decreases with $r$ for
the HFS. 

These results can also be deduced from the approximate solutions
given in Sects.~\ref{Sec:Fre} and \ref{Sec:Landau}. In particular,
taking the derivative of (\ref{Eq:barwlandau}) with respect to $u$ we see that
$\partial \bar w/\partial u < 0$ always. Since $u$ is an increasing
(decreasing) function of $r$ for the HFS (LFS), it follows that
fitness decreases with $r$ in the former case but increases in the
latter. For $r \to \infty$ the solutions (\ref{Eq:minusD},\ref{Eq:Landau_LFS}) approach
$u_\mathrm{HFS} \to 1$ and $u_\mathrm{LFS} \to -1$, respectively, with
corresponding limiting fitness values $\bar w_\mathrm{HFS} \to  w_3(1-2 \mu)$
and $\bar w_\mathrm{LFS} \to  w_0(1-2 \mu)$. As can be anticipated from
Eq.~(\ref{Eq:03}) (see also discussion in Sect.~\ref{Sec:gen}), the limit
is approached from above for the HFS but from below for the LFS. Note
that in the symmetric case ($t = 0$) the fitness is $\bar w = w_3 (1-2
\mu) = w_0 (1-2\mu)$ independent of $r$ for $r > r_c$ 
(see Appendix~\ref{Sec:Sympeak}). 

\subsection{\label{Sec:asymmetric}Asymmetric valley fitnesses}

In this final subsection, we will consider briefly how the results would
be affected if $w_1 =1-t-s_1$ and $w_2 = 1 - t - s_2$ with $s_1 \neq s_2$ 
(without loss of generality, we can set $s_1< s_2$). 
We first show that bistability requires reciprocal sign epistasis, i.e. both $s_1$ and $s_2$ have to be positive.
To see this, suppose that $s_1 < 0$ and $s_2 > 0$, such that the ordering of the fitness values is 
$w_2 < w_0 < w_1 < w_3$. Then positivity of Eq.~(\ref{Eq:12}) requires $\bar w > (1- 2 \mu) w_1$ or 
$\bar w < (1 - 2 \mu) w_2 < w_2$. The latter possibility is ruled out because the mean fitness cannot
be lower than the fitness of the least fit genotype, and the former inequality contradicts the 
condition $\bar w < (1 - 2 \mu) w_0$ imposed on the LFS by the positivity of Eq.~(\ref{Eq:03}). 
We conclude that only the HFS with mean fitness in the range $(1 - 2 \mu) w_3 < \bar w < w_3$ can exist.

To extract some information about the case $s_2 > s_1 > 0$, 
we introduce the variable $u$ in a similar way to Eq.~(\ref{Eq:defu}) such that
\begin{equation}
\label{Eq:u_gen}
f_0 = ( 1 - f_1 - f_2) \frac{1-u}{2},\quad f_3 = (1 - f_1 - f_2) \frac{1+u}{2},
\end{equation}
which yields again Eq.~(\ref{Eq:barwlandau}) for $\bar w$.
The above parametrization along with Eq.~(\ref{Eq:12}) gives
\begin{equation}
f_1 = f_2 + \frac{2 u (s_2 - s_1)}{t (1+u) + 2 u s_1} f_2.
\end{equation}
Thus, if $s_2 - s_1 \ll t, s_1, s_2$, setting $f_1 = f_2 = f$ is 
not a bad approximation and the results presented above remain valid.

When $s_2 - s_1$ is comparable to the other parameters the calculation
is much more complicated. We will not treat this case in any detail, 
but we can provide necessary conditions for bistability. First, the necessary condition for 
$r$ given in Eq.~(\ref{Eq:r_cond}) is still valid, because 
Eq.~(\ref{Eq:r_cond}) is obtained only from the denominator of 
Eq.~(\ref{Eq:link}). A necessary condition on $\mu$ similar to Eq.~(\ref{Eq:mu_cond}) 
is also available from the requirement that $w_2 < \bar w < (1-2 \mu) w_0$, 
but the result is not very useful because it is independent of the magnitude (or sign) of $s_1$.
To get a refined condition, we need some futher analysis.

From the definition of $\bar w$, we find 
(see also the Mathematica file in online supplement)
\begin{eqnarray}
\label{Eq:f1f2}
f_2+f_1 =\frac{\left(t \left(2 \mu  (1-u)+u^2-1\right)+4 \mu  u\right) (u
    (s_1+s_2)+t (u+1))}{u \left(s_1 \left(4 s_2 u+t
    (u+1)^2\right)+t (u+1)^2 (s_2+t)\right)},
\end{eqnarray}
which should be smaller than $1$ for $-1 < u <1$. Now assume that a critical value 
$r_c$ exists ($r_c <r \le 1$) beyond which bistability appears.
For $r=1$ Eq.~(\ref{Eq:link}) shows that
$f_0 f_3 = f_1 f_2$ for any equilibrium solution. On the other hand, we 
expect that for small $\mu$ the frequencies of both valley genotypes are small, 
$f_1 f_2 \ll 1$. Computing $f_0 f_3$ using Eq.~(\ref{Eq:u_gen}) this 
is seen to imply that $u$ for the LFS should be very close to $-1$ when 
$r=1$. Expanding Eq.~(\ref{Eq:f1f2}) around $u = -1$ gives 
\begin{equation}
f_1+f_2 = \mu \frac{(s_1 + s_2)(1-t)}{s_1 s_2} + \frac{t}{2 s_1 s_2} (s_1 + s_2 - \mu (2+ s_1 + s_2 - 2 t)) ( 1 + u) + O(1+u)^2.
\end{equation}
For a LFS to be possible, the leading order term should be
smaller than unity, which gives 
\begin{equation}
\mu < \frac{s_1 s_2}{(s_1+s_2) (1-t)} = \frac{s_H}{2(1-t)},
\label{Eq:diffw1w2}
\end{equation}
where $s_H$ is the harmonic mean of $s_1$ and $s_2$. Note that
Eq.~(\ref{Eq:diffw1w2}) reduces to Eq.~(\ref{Eq:mu_cond}) when $s_1=s_2$.
Hence, if $0 < s_1\ll s_2$, $\mu$ must be smaller
than $s_1/(1-t)$ to have multiple solutions.

\section{\label{Sec:Dis}Discussion} 

In this work we have presented a detailed analysis of a deterministic,
haploid two-locus model with a fitness landscape displaying reciprocal
sign epistasis. We have established the conditions for the occurrence
of bistability, and derived accurate approximations for the critical
recombination rate $r_c$ at which bistability sets in. 
For $r < r_c$ there is a single equilibrium solution in which the fittest genotype 
is most populated. For $r > r_c$ we find two stable equilibrium solutions, one of 
which is concentrated on the fittest genotype (the HFS) and a second one concentrated
on the lower fitness peak (the LFS). For $\mu \to 0$ these two solutions become point
measures, in the sense that $f_3 \to 1$ and $f_0 \to 1$, respectively, but for any finite
mutation probability all genotypes are present at nonzero frequency. 

We briefly summarize the most imporant quantitative results presented in this paper. 
The expressions (\ref{Eq:rc_mu}) and (\ref{Eq:rc_t}) for $r_c$ are based on a systematic expansion
in terms of the mutation probability $\mu$ and the peak
asymmetry $t$, respectively, while the interpolation formulae
Eq.~(\ref{Eq:rc_match}) and Eq.~(\ref{Eq:rc_t_match}) provide numerically accurate values of $r_c$
over a wide range of parameters. In particular, our results show that
the lower bound (\ref{Eq:r_cond}) on $r_c$ becomes an equality for $\mu \to 0$,
which is consistent with earlier results obtained either directly 
at $\mu = 0$ \citep{Feldman1971,Rutschman1994} or using a
unidirectional mutation scheme \citep{Eshel1970,Karlin1971}. Clearly the limiting
behavior for $\mu \to 0$ should not depend on the mutation scheme
employed. 
Approximate results for the stationary frequency
distributions are found in
Eqs.~(\ref{Eq:small_mu_f0}), (\ref{Eq:small_mu_f3}) and 
Eqs.~(\ref{Eq:f1t}), (\ref{Eq:f3t}), (\ref{Eq:f0t}).
Of particular interest are the simple
formulae derived from the cubic equation (\ref{Eq:Landau}), which are
remarkably accurate close to the bistability threshold and for small
$t$.

The consequences of our results for the question of a possible evolutionary advantage of recombination 
are twofold. Dynamically, the onset of bistability implies that recombination 
strongly suppresses the escape of large populations from suboptimal fitness peaks. In the deterministic
infinite population limit considered here, the escape time diverges at $r = r_c$ 
\citep{Jain2010}, whereas in finite populations one expects a rapid
(exponential) increase of the escape time for $r > r_c$
\citep{Stephan1996,H1998,Weinreich2005a}. 
In a multipeaked fitness landscape, recombination can 
therefore dramatically slow down adaptation \citep{deVisser2009}. On the other hand, we have seen
in Sect.~\ref{Sec:fitness} that the equilibrium mean fitness may increase or decrease with the recombination rate
when $r > r_c$ depending on which of the two equilibria is considered. In a fitness landscape with more than two peaks one 
anticipates an even richer structure of stationary solutions with a correspondingly complex dependence
on recombination rate.  

It would be of considerable interest to extend the present study to 
finite populations. 
For the case of symmetric peaks ($t=0$)
this problem has been addressed by \citet{H1998} in the framework of 
a diffusion approximation. A key step in his analysis was the
reduction to a one-dimensional problem by fixing the frequency of the
valley genotypes at its stationary value $f = \mu/s$. 
However, we have seen above that in the case when $s$ and $t$ 
are large in comparison to $\mu$, $f$ cannot be treated as a constant
and therefore the reduction to a one-dimensional diffusion
equation is generally not possible. Some progress could be made in the
regime where the (effectively one-dimensional) Landau equation 
(\ref{Eq:Landau}) applies, and we intend to pursue this approach in
the future.

\section*{Acknowledgments}

We wish to thank to Alexander Altland, Reinhard B\"urger, Arjan de
Visser, Paul Higgs and Kavita Jain for useful discussions.
Support by Deutsche Forschungsgemeinschaft within SFB 680
\textit{Molecular Basis of Evolutionary Innovations}
is gratefully acknowledged.
In addition,  J.K. acknowledges support
by NSF under grant PHY05-51164 during a visit at KITP, Santa Barbara,
where this work was begun, and S.-C.P. acknowledges support by
the Catholic University of Korea, Research Fund, 2010.    

\appendix
\section{\label{Sec:Hx}The function $h(x)$ and its values at selected points}
All information in this appendix can be found in the 
Mathematica file in  online supplement, and is provided
here only for completeness.

By squaring both sides of Eq.~(\ref{Eq:eqbarw}),
the steady state mean fitness becomes the solution
of the equation $h(x) = 0$
where
\begin{eqnarray}
4(\bar w-w_1)^2 - \frac{B}{A} (w_3 + w_0 A - (1+A) \bar w)^2
=\frac{4 (1-t-x)h(x)}{x (t+x)( (1-t-x)^2 - (1-r)(1-t))} .
\end{eqnarray}
As shown in the Mathematica file in online supplement, 
$h(x)$ takes the form $h_0(x) + r h_1(x)$ with
\begin{eqnarray}
h_0(x) &=& b_4 x^4 + b_3 x^3 + b_2 x^2 +  b_1 x - b_0,\\
h_1(x) &=& -(1-2\mu)^2 c_3 x^3 - (1-2\mu) c_2 x^2 +  c_1 x - c_0,
\end{eqnarray}
where
\begin{eqnarray}
b_4 &=&  (1-2\mu)(2s + t) ,\quad
c_3 =  2s + t - (s+t)^2 ,\nonumber\\
b_3 &=&  (t^2 + 2 s t - 2 s^2)(1-2 \mu) + \mu^2 ( 4 c_3 + t^2 ),\quad
c_2 = (t + 2 \mu - 4t \mu) c_3 - s^2,\nonumber\\
b_2 &=& -3 s^2 t ( 1- 2 \mu) - \mu^2 \left [ 4 (1-2 t)c_3 + 3 t^2(1-t)\right ],\nonumber\\
c_1 &=& t (1-2\mu) (s^2 - 2 \mu (1-t) c_3 ) + \mu^2 t^2 (1-s-t)^2,\nonumber\\
b_1 &=&-(1-2\mu) s^2 t^2 - \mu^2 t \left [ (4-5 t) c_3 - t (1-t)(2-3 t)
\right ],\nonumber\\
c_0 &=& (1-t) (1-s-t)^2t^2 \mu ^2 ,\quad
b_0 =  (1-t) (2-s-2 t)s t^2 \mu ^2.
\end{eqnarray}
Note that $b_4>0$ if $\mu<\frac{1}{2}$.
The values of $h(x)$ 
at $x=0$, $x=-t$, and $x=x_1\equiv w_0 - w_1/(1-2\mu)$ are
\begin{eqnarray}
h(0) &=& - b_0 - r c_0 ,\\
h(-t) &=& - t^2 \mu^2 \left [ 1 - \left (1-r \right )(1-s-t)^2  \right ]  ,\\
h(x_1) &=& - \frac{w_1}{(1-2\mu)^3} \left [1- \left (1-r \right )
(1-2\mu)^2 \right ] (s(s+t)(1-\mu) - c_3 \mu )^2,
\end{eqnarray}
which are all negative if $0<\mu<\frac{1}{2}$.
\section{\label{Sec:Sympeak}Solution for symmetric fitness peaks ($t=0$)}
When $t=0$, our problem is reduced to that (approximately) solved by 
\citet{H1998}. For the paper to be self-contained, we solve the case with
$t=0$ exactly in this appendix. 

When $t=0$, Eq.~(\ref{Eq:03}) suggests 
either $f_3 = f_0$ or $\bar w = 1-2 \mu$. Let us first consider 
$\bar w = 1-2\mu$.
Needless to say, this solution is impossible if $\mu \ge \frac{1}{2}$.
Since $\bar w = (f_0 + f_3) + 2 f (1-s) = 1 - 2 f s$,
we get
\begin{equation}
f =\frac{\mu}{s},\; f_0 + f_3 = 1- 2 \frac{\mu}{s}.
\end{equation}
From Eq.~(\ref{Eq:link}) with $\bar w = 1-2\mu$, 
we get
\begin{eqnarray}
(f_0 - f_3)^2 &=& (f_0 + f_3)^2 - 4 f_0 f_3 = \frac{2}{r s}(1-2\mu)\xi,
\end{eqnarray}
where $\xi = (2-s) (\mu_{c0}-\mu)(r-r_{c0})$ with $r_{c0}$ and
$\mu_{c0}$ given in Eqs.(\ref{Eq:muc0}) and (\ref{Eq:rc0}).

To have an asymmetric solution ($f_0 \neq f_3$), $\xi$ should be nonnegative.
Because $\mu > \mu_{c0}$ implies $r_{c0} <0$, $\xi$ is nonnegative
only if $r\ge r_{c0}$ and $\mu < \mu_{c0}$ (note that when $\mu=\mu_{c0}$,
$\xi$ is nonzero and negative, because Eq.~(\ref{Eq:rc0}) diverges as 
$1/(\mu_{c0} - \mu)$ for $\mu \to \mu_{c0}$).
Since $r_{c0}$ cannot
be larger than 1,
the more restrictive condition on $\mu$ becomes
$\mu < \mu_c$ where $\mu_c$ is the solution of the equation
$r_{c0}=1$ given by
\begin{equation}
\mu_c = \frac{s}{4}.
\end{equation}
Note that $\mu_c$ is the same as $\mu_{c0}$ up to leading order of $s$.

Now let us find the solution with $f_3 = f_0$. 
From $1 - 2 f s = \bar w$ and
Eq.~(\ref{Eq:link}) along with the substitution $\bar w = (1-2\mu) (1+y)$,
we obtain
\begin{equation}
g_s(y)\equiv   y^2 +  \left ( r(1-s) + s + \xi \right ) y + \xi =0.
\label{Eq:gsy}
\end{equation}
Since
\begin{eqnarray*}
g_s(-s) = -(1-s)(2-s)\left [\frac{2 \mu^2}{1-2\mu} +  r( \mu  + \mu_{c0})\right ]
\end{eqnarray*}
is negative,
there is only one solution with $y > -s$ which is
\begin{equation}
y = \frac{-2 \xi}{ r(1-s)+s + \xi + 
\left ( (r (1-s) + s + \xi)^2 - 4 \xi \right )^{1/2} }.
\end{equation}
When $\xi < 0$ (either $\mu\ge \mu_{c0}$ or $\mu<\mu_{c0}$ together with $r < r_{c0}$), 
the larger  solution is nonnegative.
On the other hand, if $\xi>0$ ($\mu<\mu_{c0}$ and 
$r>r_{c0}$), the larger solution which is still larger than $-s$ is negative.

\bibliographystyle{spbasic}      
\bibliography{me}   

\end{document}